# A hybrid method for overcoming thermal shock of non-contact infrared thermometers

**Yaonan. Tong,**[a,1] **Shunnan. Zhao,**[a,b] **, Haitao. Yang,**[a] **Zhiqi. Cao,**[a] **Song. Chen,**[a] **and Zhenguang. Chen,**[b]

[a] *School of Information Science and Engineering, Hunan Institute of Science and Technology,*
  *439 Xiangbei Avenue, Yueyang City, China*

[b] *Dongguan Zhenhai Electronic Technology Co., Ltd,*
  *6 Minfu Road, Liaobu Town, Dongguan City, China*
  *E-mail*: tongyaon@hnist.edu.cn

**ABSTRACT**: The non-contact infrared thermometer (NCIT) is an important basic tool for fever screening and self-health monitoring. However, it is susceptible to the thermal shock when working in a low temperature environment, which will cause a time-consuming and inaccurate human body temperature measurements. To overcome the effects of thermal shock, a hybrid temperature compensation method combining hardware and algorithm is proposed. Firstly, the principle of infrared temperature measurement is described and the influence of thermal shock on infrared thermometer is analyzed. Then, the hybrid temperature compensation scheme is constructed by mounting a heating ring on the infrared sensor shell, and using the proportional integral derivative (PID) algorithm and the pulse width modulation (PWM) technology to control it heating. In this way, the internal ambient temperature of infrared sensor can be raised rapidly closing to the external ambient temperature, and the stable outputs of the infrared sensor are also accelerated. Finally, some experiments are carried out in a laboratory. The results show that the proposed method can quickly and accurately measure the temperature of standard black body when the ambient temperatures are 5 ℃, 15 ℃ and 25 ℃ respectively, the measurement error is only ± 0.2 ℃, and the measurement time is less than 2 seconds. This study would be beneficial to improve performance of NCIT, especially the infrared ear thermometer.

**KEYWORDS**: Thermal shock; Non-contact infrared thermometer (NCIT); Proportional integral derivative (PID) algorithm; pulse width modulation (PWM)

---

[1] Corresponding author.

# Contents

## 1. Introduction

The infrared temperature measurement technology has been widely used in industrial, medical, civil, military and other fields due to its many advantages, especially fast measurement speed, convenient use, and no physical contact with the measured objects. In the medical fields, several global or local epidemics in recent years, such as Severe Acute Respiratory Syndrome (SARS), Ebola virus, Swine influenza, and COVID-19, have accelerated the development and application of fever screening technology [1-5]. The thermometer for fever screening should not only be cheap, simple, noninvasive, fast and safe, but also be accurate, reliable and reproducible [2]. Commonly used fever screening technologies include contact temperature measurement tools, typically mercury-in-glass thermometer, and non-contact temperature measurement tools, such as infrared thermal imaging scanner [6-8] and infrared thermometer [9-12]. As a traditional clinical human body temperature measurement tool, the mercury-in-glass thermometer has high reliability and wide application. But because the working principle of mercury thermometer is metal expand on heating and contract on cooling, it needs a long time (in minutes) of full contact with the measured object (e.g., armpits rectum or oral cavity) to accurately measure the body temperature, and the reading is not convenient [2]. What's worse, mercury is a poisonous substance, which may bring hidden danger to the operator or the measured people. As an alternative, the infrared thermometers have been widely used. The infrared thermal imaging scanners have been used to screen suspected fever cases in recent years, especially in the period of COVID-19 pandemic, and have the advantage of measuring multiple targets at the same time [2,7]. However, their accuracy and reliability of human body temperature measurement are not high enough and the cost is relatively expensive, so it is difficult to meet the low-cost and highly reliable requirements for epidemic prevention and control.

The non-contact infrared thermometer (NCIT) has become an important basic tool for the government to screen fever-type epidemics and personal self-health monitoring due to its low cost, fast measurement speed and convenient operation[1-5]. According to Planck radiation law, any object whose temperature is higher than absolute zero is emitting infrared radiation to the surrounding environment, and the radiation energy is directly related to the surface temperature of the object. Based on this principle, the NCIT uses an infrared sensor to absorb the infrared energy radiated by the human body and convert it into an electrical signal, so as to calculate the body temperature by micro-controller unit (MCU) or digital signal processor (DSP) [1, 2, 9]. The NCIT is usually hand-held, and its measurement site is the forehead, temples, neck artery and tympanic cavity of the human body. The hand-held NCIT performance depends on the operator and the measure distance. To solve the problem of operation dependence, Tomy Abuzairi et al. [9] developed an infrared forehead thermometer which can be mounted on the wall and automatically measure human body temperature when the distance between the sensor and forehead is adequate.

The NCIT measurement accuracy is a widely concerned problem [3, 10-13]. Recently, Chen et al. [3] conducted a very interesting comparative study on 45 rehabilitation inpatients (the average age was 54.71 ± 17.73 years, 28 men) by using three kinds of temperature measuring instruments, that are NCIT, mercury axillary thermometer (MAT) and infrared tympanic thermometer (ITT). They concluded that NCIT is superior to ITT and MAT in measuring the body temperature clinically, and the NCIT can save measurement time and effectively avoid unnecessary pain of patients. In paper [10], a cross-sectional study was done by recording axillary temperatures of 250 febrile infants (the age from 1st day to 12 month) over a period of two months using NCIT and mercury thermometer, respectively. The study results show that the emergent digital infrared thermometry can be compared favorably with the traditional mercury thermometer in terms of measurement accuracy, the mean difference is only 0.016. But in [12], a study was carried out on 401 children (interquartile range 0.79-3.38 years, 203 boys) attending with an acute illness by using NCITs, electronic axillary and infrared tympanic thermometers, and the conclusion is that the 95% limits of agreement are > 1 ℃ for both NCITs compared with electronic axillary and infrared tympanic thermometers, which could lead to wrong clinical decisions.

Therefore, how to effectively improve the accuracy and reliability of NCIT equipment is a very valuable research topic [14-21]. Many scholars and engineers have adopted different methods to solve these problem, specially the ambient temperature compensation, which can be divided into hardware and algorithm schemes. Shen et al [14] proposed a novel dual-mode modulation infrared temperature sensing method based on a switching circuit composed of an operational amplifier AD8551 and a p-channel MOSFET, therefore, only a single thermopile sensor is needed to realize infrared radiation and ambient temperature sensing. After ambient temperature compensation, the error of infrared temperature measurement is less than 0.14 °C. Lü et al [15] developed a dual-band infrared thermometer considering both reflected radiation from the measured target and stray radiation inside the measurement system, so there is no need to repeat calibration when the ambient temperature fluctuates. In [16], a chopper stabilized operational amplifier is used instead of a precision operational amplifier to overcome the temperature-induced zero offset drift, which reduces the root mean square (RMS) noise of infrared radiation thermometer from 5K to 1K. On the other hand, there are many valuable researches on the ambient temperature compensation algorithm for infrared thermometer, such as infrared sensor signal polynomial compensation algorithm [17], incidence angle compensation algorithm [18], distance compensation algorithm [20], uncertainty components compensation function [20] and infrared focal plane non-uniformity correction algorithm [21]. The above methods based on hardware or algorithm have played an important role in improving the NCIT performance. However, we often ignore a special but important problem, that is, the NCIT is susceptible to thermal shock or even does not work under extremely low ambient temperature, which often occurs in low temperature winter or cold zone areas.

The thermal shock is an important topic in the study of material properties [22], but the phenomenon and harmful affects of thermal shock in NCIT have not been paid much attention. As a matter of fact, the thermopile infrared sensor is easily affected by the thermal shock, especially for infrared ear thermometer when working in extremely low temperature environments. At this time, the temperature in the inner cavity of the infrared sensor is very low, but the temperature of the ear canal is much higher, which close to the temperature of the human body. Therefore, the ear canal will suddenly and greatly radiate heat into the inside of the sensor, and the exciting ambient temperature compensation algorithms are difficult to play an effective role, which will cause a very large measurement error of the infrared thermometer.

In this paper, a hybrid ambient temperature compensation method based on hardware and PID control algorithm is proposed to preheat the infrared sensor, so that the NCIT may not be affected by external environment factors including thermal shock.

## 2. Principle and thermal shock of infrared thermometer

### 2.1 Principle of infrared thermometer

The working principle of infrared thermometer is the Planck's law of black body radiation: When the temperature of an object is higher than absolute zero, it will continuously emit radiation energy to the surrounding. The radiation emissivity $M$ of a spectral wavelength $\lambda$ at a certain absolute temperature $T$ is shown as follow:

$$M(\lambda, T) = \frac{c_1}{\lambda^5}[e^{\frac{c_2}{\lambda T}} - 1]^{-1} \tag{2.1}$$

Where $c_1$ and $c_2$ are the first and second Planck radiation constants respectively. The total radiation emissivity of all spectral wavelengths of a black body at a certain absolute temperature $T$ can be obtained by integrating formula (2.1):

$$M(T) = \int_0^\infty \frac{c_1}{\lambda^5}[e^{\frac{c_2}{\lambda T}} - 1]^{-1} d\lambda = \sigma T^4 \tag{2.2}$$

Where $\sigma$ is Stefan Boltzmann constant, and $\sigma = 5.67 \times 10^{-8} W/(m^2 \cdot K^4)$.

It can be seen from equation (2.2) that the total radiation power of black body is proportional to the fourth order of its thermodynamic absolute temperature $T$. The infrared sensors installed in NCIT selectivity collect radiation energy in a certain infrared wavelength range (usually from 9 μm to 13 μm), so the upper and lower limits of integration in equation (2.2) should be replaced by the selected wavelength range.

Because the measured object is easily affected by the environments, its radiation power is always less than that of the black body under the same conditions. Let the object temperature be $T_{obj}$, the ambient temperature be $T_{env}$, the radiation energy emitted by the object surface be $\varepsilon \sigma T_{obj}^4$, the radiation energy absorbed be $\alpha \sigma T_{env}^4$, so the actual radiation energy of the object is:

$$J = \varepsilon\sigma T_{obj}^4 - \alpha\sigma T_{env}^4 \qquad (2.3)$$

Where $\varepsilon$ is the emissivity and $\alpha$ is the absorptivity of the object. If $\varepsilon$ and $\alpha$ is equal, it can be obtained from equation (2.3) that [23]:

$$J = \varepsilon\sigma(T_{obj}^4 - T_{env}^4) \qquad (2.4)$$

According to equation (2.4), if the radiation energy $J$ and the ambient temperature $T_{env}$ are known, the absolute temperature of the measured object $T_{obj}$ can be calculated. This is the basic calculation formula of infrared temperature measurement.

A typical structure of infrared temperature sensor thermopile based is shown in figure 1. In the figure, the thermopile, designed based on Seebeck thermoelectric effect, senses the object radiation energy $J$ and usually converts it to a millivolt voltage for output. The reference sensor, usually composed of negative temperature coefficient (NTC) thermistor, is used to measure the environment temperature $T_{env}$, which may be used for ambient compensation algorithms.

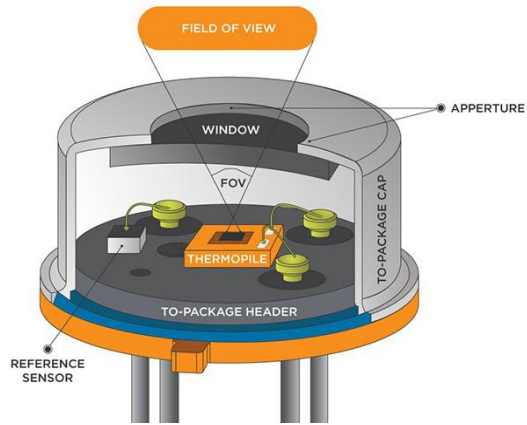

**Figure 1.** Structure of infrared temperature sensor based on thermopile. Cited from TE connectivity company, https://www.te.com/usa-en/product-CAT-IR0001.html.

### 2.2 Thermal shock and its influence

The thermal shock refers to a large amount of heat exchange in a short time due to rapid heating or cooling. When the temperature changes dramatically, the material will produce impact thermal stress, and the devices made of it may work abnormally. For infrared thermometer, if the working environment temperature changes suddenly, it usually means from low temperature environment to high temperature environment, the thermal shock will affect its measurement. For an example, when the infrared ear thermometer in low temperature environment is inserted into the ear canal with higher environment temperature, the heat of the ear canal will transfer into the infrared thermometer, and the internal thermopile and NTC thermistor of the sensor will be affected seriously.

**(a) Influence of thermal shock on thermopile of infrared sensor**

Because the thermopile is composed of several thermocouples, so the effect of thermal shock on thermopile can be analyzed by the step temperature response of thermocouple, which can be described as follow [24]:

$$T - T_0 = (T_e - T_0)(1 - e^{-t/\tau}) \qquad (2.5)$$

In equation (2.5), $T$ is the indicated temperature of the thermocouple; $T_0$ is the initial temperature of the thermal contact point; $T_e$ is the step temperature; $\tau$ is the time constant of thermocouple, which refers to the required time for the difference between the measured temperature $T$ and the initial temperature $T_0$ of the thermocouple to reach 63.2% of the temperature step ($T_e - T_0$). Because the time constant $\tau$ is different for different thermocouples, the response speed is different. Typically, the step temperature response curve is shown in figure 2.

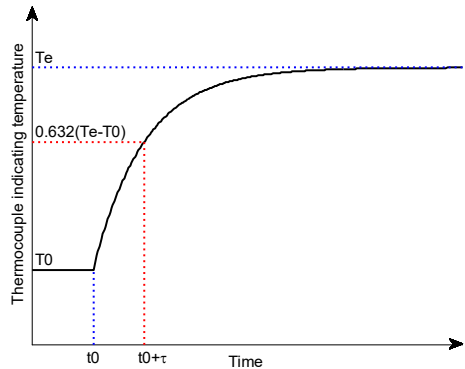

**Figure 2.** Step temperature response curve of thermocouple.

Because the thermopile consists of several thermocouples, the thermal shock response characteristic of the thermopile-based infrared sensor is similar to the step temperature response of the thermocouple. According to the figure 2, if the initial temperature $T_0$ of the thermal contact point of the thermopile in the infrared sensor is far lower than the object temperature $T_e$, the response time of the thermopile will be longer, that is, it will take a longer time for NCIT to accurately perceive the actual temperature of the human body.

**(b) Influence of thermal shock on reference sensor**

The reference sensor of the infrared sensor is usually made by NTC thermistor. The thermal shock of infrared thermometer also will cause the reading of NTC reference sensor unstable before the internal thermal balance of sensor. The NTC reading reflects the ambient temperature of ear canal, so that the ambient compensation algorithm will not work correctly.

According to the above analysis, it is difficult to accurately calculate the absolute temperature of the measured target according to formula (2.4), especially in the initial stage of infrared thermometer measurement. Therefore, if the internal ambient temperature of the sensor can be preheated to a value closing to the external in advance, the influence of the thermal shock may be effectively avoided, and the speed and accuracy of temperature measurement can be improved accordingly.

## 3. Sensor preheating scheme and control method

In order to reduce the influence of environmental temperature on infrared thermometer, various ambient compensation algorithms are proposed by researchers, but the thermal shock effect cannot be effectively eliminated by these algorithms, specially for infrared ear thermometer. Therefore, a solution of preheating the infrared sensor was proposed to make its internal temperature close to the external one before formal measurement, so as to reduce or eliminate the effects of thermal shock.

### 3.1 Hardware scheme of sensor preheating

The details of the proposed system structure and sensor preheating improvement scheme are shown in figure 3 and figure 4, respectively. As we can see from figure 3 that this improved scheme only be added a heating material and its control circuit based on the existing NCIT structure. The scheme diagram consists of an operational amplifier for amplifying thermopile output voltage, two analog-to-digital converters (ADC) for detecting the thermopile and NTC outputs, a micro-controller unit (MCU) and display module. The MCU records the infrared sensor's internal ambient temperature detected by the NTC, and uses PID algorithm combined with PWM wave to feedback control the heating, further to keep the internal ambient temperature of the infrared sensor at a suitable value.

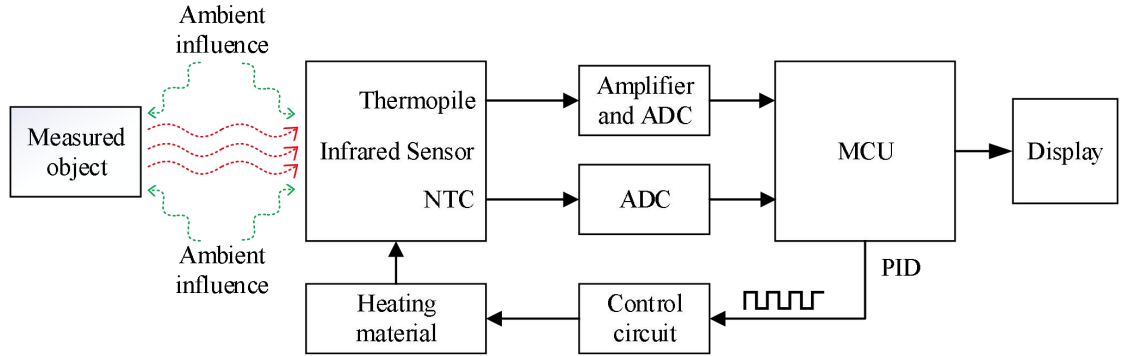

**Figure 3.** NCIT preheating and control block diagram.

Figure 4 shows the diagram of sensor preheating material installation and control circuit schematic. The left and right parts in figure 4(a) are the prototype appearance of the common infrared sensor and the new structure after improvement. An electric heating ring is mounted on the infrared sensor shell, and two pins, named "heat +" and "heat -" as shown in the figure, are led out for connecting heating control switch circuit. The heating ring is adhered to the sensor with thermal conductive silica gel, so it is electrically insulated from the sensor but has good heat conduction performance. The figure 4(b) is the PWM switch circuit for the heating control. One end of the circuit is connected to I/O port of MCU, receiving PWM wave control signal, and the other end (Heat +5V) is connected to the positive pole of the heating ring.

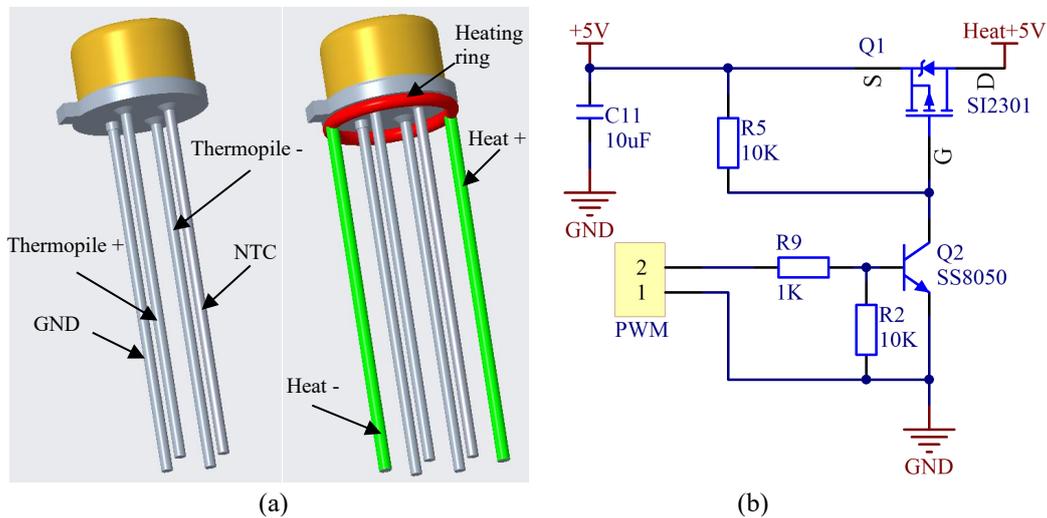

(a)        (b)

**Figure 4.** Infrared sensor preheating modification details. (a) The heating ring is mounted on the metal shell of the sensor. (b) The preheating control switch circuit.

### 3.2 Preheating PID control strategy and algorithm

This paper uses digital PID algorithm to control the ambient temperature inside the infrared sensor by preheating method. The control strategy can be described as follows. (i) The NTC

– 6 –

thermistor in the infrared sensor is used to detect the internal actual ambient temperature, which is converted into a digital signal by ADC and stored in the MCU memory. (ii) The temperature deviation is obtained by comparing the collected actual ambient temperature with the set object ambient temperature value. (iii) According to the ambient temperature deviation, PID algorithm is utilized to output the PWM regulation wave with corresponding duty cycle. Then the heating power is controlled by the PWM wave, so as to make the sensor internal ambient temperature gradually close to the set temperature value. (iv) For the infrared ear thermometer, it is a good choice to set the heating target temperature at 33 °C, which is close to the ear canal actual environment. (v) If the actual ambient temperature is higher than the set temperature, the heating should be stopped.

The PID control method is widely used in industrial field, and it is the most simple and effective algorithm in the feedback algorithm. Compared with other complex algorithms, we use digital PID algorithm to reduce the calculation cost of infrared thermometer. The PID control algorithm in continuous time domain is described as follow:

$$u(t) = K_p \left[ e(t) + \frac{1}{T_1} \int_0^t e(t) dt + T_2 \frac{de(t)}{dt} \right] \quad (3.1)$$

In formula (3.1), $u(t)$ is the output of the PID control algorithm, used to control the preheating; $e(t)$ is the deviation value, that is the difference between the actual value of the NTC thermistor and the set target ambient temperature value; $K_p$ is the proportional parameter, $T_1$ is the integral parameter, and $T_2$ is the differential parameter. Each item in formula (3.1) can be discretized as following formula:

$$\begin{cases} \int_0^t e(t) dt = T_c \sum_{i=0}^{n} e(iT_c) = T_c \sum_{i=0}^{n} e(i) \\ \frac{de(t)}{dt} = \frac{e(nT_c) - e[(n-1)T_c]}{T_c} = \frac{e(n) - e(n-1)}{T_c} \end{cases} \quad (3.2)$$

Where $T_c$ is the sampling time and $n$ is the sampling sequence number.

By introducing formula (3.2) into equation (3.1), the expression of discrete equation of digital PID control algorithm can be obtained when $t = n$:

$$u(n) = K_p \left[ e(n) + \frac{T_c}{T_1} \sum_{i=0}^{n} e(i) + T_2 \frac{e(n) - e(n-1)}{T_c} \right] \quad (3.3)$$

By simplifying equation (3.3), we can get the following results:

$$u(n) = K_p e(n) + K_1 \sum_{i=0}^{n} e(i) + K_2 \frac{e(n) - e(n-1)}{T_c} \quad (3.4)$$

Where $K_1 = K_p \times \frac{T_c}{T_1}$, which is the integral parameter, and $K_2 = K_p \times T_2$, which is the differential parameter.

## 4. Measurements

### 4.1 Materials and apparatus

To verify the effectiveness of the proposed hybrid method for overcoming the thermal shock influence of the infrared thermometer, we have built an experimental system as shown in figure 5. The materials and apparatus include specially adapted infrared sensor, sensor signal acquisition and preheating control circuit board as shown in figure 5(a), self-made MCU motherboard based on STM32 chip as shown in figure 5(b), standard black body submerged in a thermostat water bath as shown in figure 5(c). All subsequent experiments were completed in a laboratory with adjustable constant temperature by air conditioner, as shown in figure 5(d).

The infrared sensor used in this experiment is STP9CF55 made by Shanghai Yeying Electronic Technology Co., Ltd. The thermostatic water bath is manufactured by Nanjing Star Experimental Equipment Co., Ltd., the product model number is GDH-0506, and the digital display accuracy is 0.001 °C. The standard black body is self-made by Dongguan Zhenhai Electronic Technology Co., Ltd., which effective emissivity is ≥ 0.999, and the effective temperature range is from 15 °C to 50 °C.

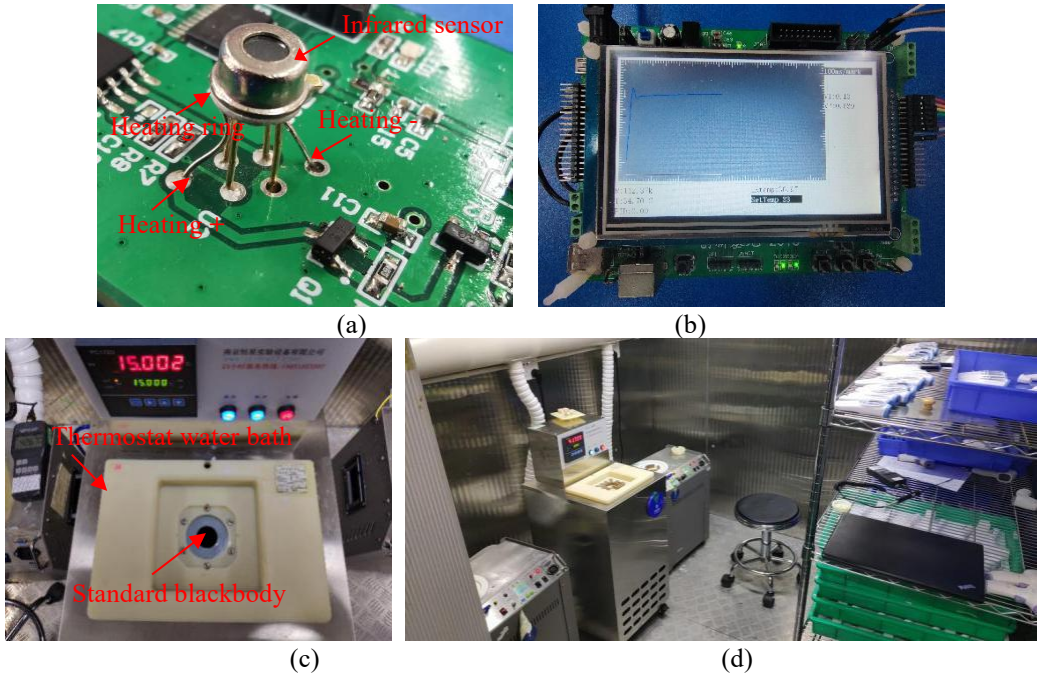

**Figure 5.** Experimental equipments and laboratory. (a) Specially adapted infrared sensor and preheating control circuit. (b) Self-made MCU motherboard. (c) Standard blackbody and thermostat water bath. (d) Constant temperature laboratory.

## 4.2 Measurement method

The black body is used as the target to be measured, and the temperature was set to 35 °C - 39 °C, increasing by 1 °C each time. The laboratory ambient temperature was adjusted by air conditioner to 5 °C, 15 °C and 25 °C, respectively. Under different laboratory ambient temperatures, a group of experiments were carried out to measure different black body temperature. Each group of experiments were done twice, the first was the direct measurement without preheating, the second was carried out under the same conditions but using the preheating method proposed in this paper. Before the measurement, the signal acquisition and preheating control circuit board with adapted infrared sensor was placed in the adjusted laboratory environment for not less than half an hour. So that the internal ambient temperature of the infrared sensor was close to the laboratory temperature.

The self-made MCU motherboard mainly played the role of reading and displaying the internal ambient temperature and the black body temperature, which were converted from the NTC resistor value and thermopile output voltage of the infrared sensor. When the experiment of the preheating method was carried out, the preheating target ambient temperature inside the infrared sensor was set at 33 °C, and the initial parameters of PID algorithm were set as $K_P = 0.7$, $K_i = 0.005$ and $K_d = 5$, respectively.

### 4.3 Results and discussion

A series of experiments were carried out at initial ambient temperature of 5 °C, 15 °C and 25 °C respectively. Firstly, a pure sensor preheating experiment was done. The preheating ambient curve was observed by detecting the NTC resistor value in the laboratory environment without the measurement of standard blackbody temperature. The infrared sensor internal ambient temperature control curves are shown in figure 6.

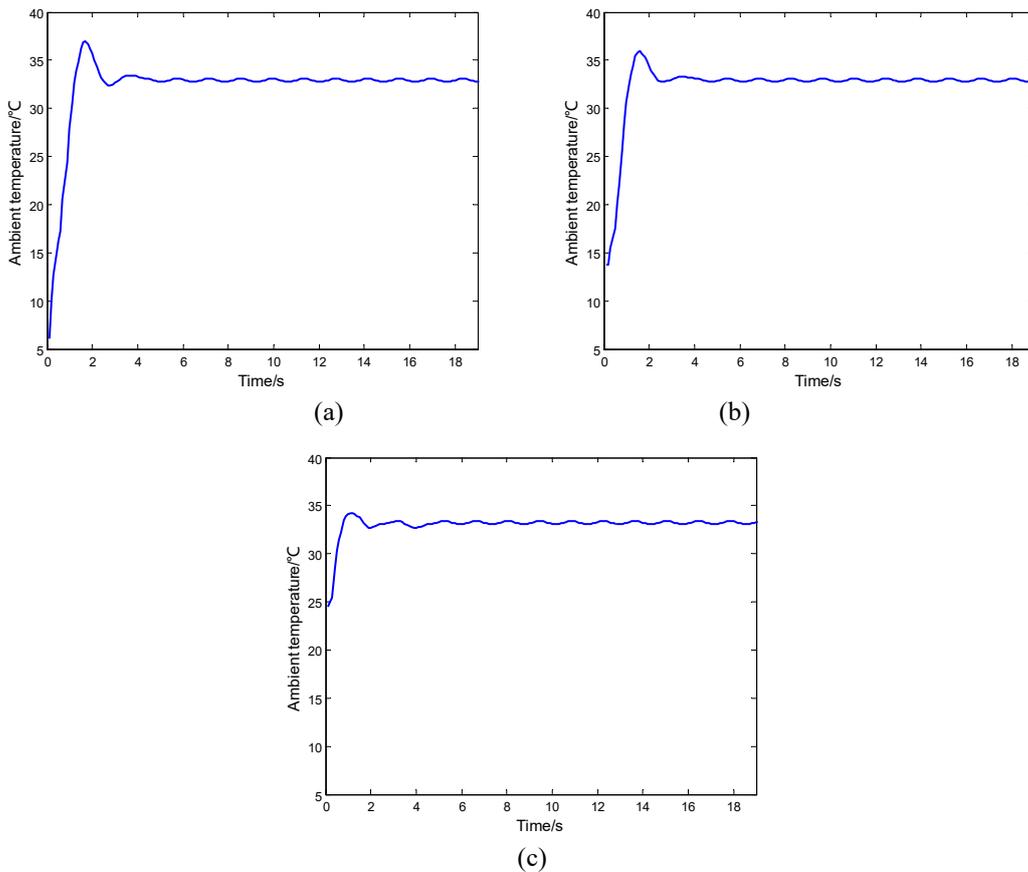

**Figure 6.** Infrared sensor internal ambient temperature curves under preheating. (a) Initial ambient temperature is 5 °C. (b) Initial ambient temperature is 15 °C. (c) Initial ambient temperature is 25 °C.

It can be seen from the figure 6 that the infrared sensor internal ambient temperature can be stabilized within about 2 seconds at different initial ambient temperatures, and as the initial ambient temperature increases, the better the PID control effect is and the overshoot and stabilization time will gradually decrease. These experimental results show the effectiveness of the preheating scheme.

Next, the temperatures of standard black body were measured occupying non heating mode and preheating mode, respectively. The infrared sensor ambient temperature and black body target temperature measure curves are shown in figure 7. In these experiments, the black body temperature and the preheating target ambient temperature inside the infrared sensor were 37 °C and 33 °C, respectively.

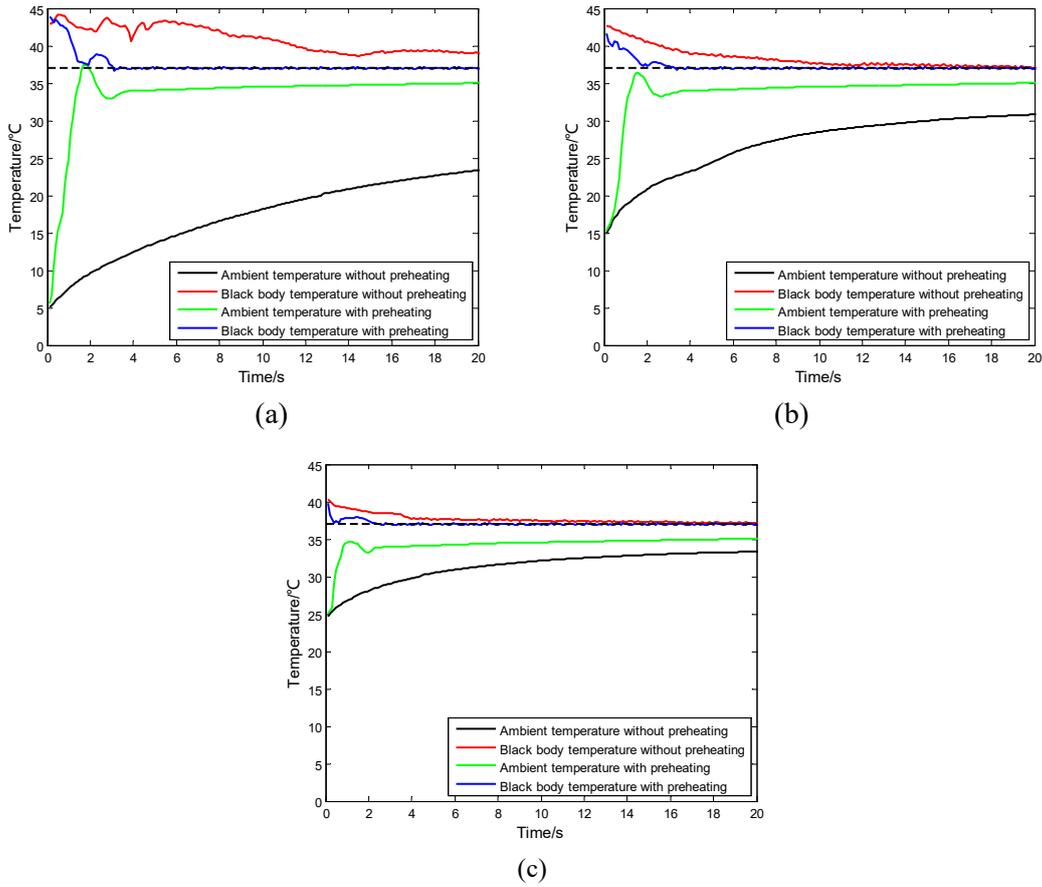

**Figure 7.** Ambient temperature and target temperature curves. (a) Initial ambient temperature is 5 °C. (b) Initial ambient temperature is 15 °C. (c) Initial ambient temperature is 25 °C.

As shown in figure 7, if the preheating method is not used, the infrared sensor needs a long time to measure the external environment temperature and the black body temperature, and the error is very large, especially in the case of low ambient temperature. In three cases, the ambient temperature and target temperature curve deviate from the actual value seriously within the first 10 seconds after the infrared sensor is inserted into the black body. The reason is that it is difficult for thermopile and NTC resistor to reach thermal balance quickly under thermal shock. Fortunately, through the preheating method proposed in this paper, the measurement time of the ambient and target temperatures is greatly shortened, and the accuracy is significantly improved. For example, under the extremely low working environment temperature of 5 ℃, the target temperature can be accurately measured about at the 3$^{rd}$ second after power on, and the ambient temperature change in the next time is small, and the impact on the target temperature measurement is also small. Therefore, the influence of thermal shock is effectively eliminated, and the speed, accuracy and stability of infrared temperature measurement have been greatly improved.

The above experiments show that when the infrared sensor internal ambient temperature is preheated to a suitable value and basically stabilized, such as 33 °C for infrared ear thermometer, then the temperature of the measured target can be calculated accurately by reading the thermopile output voltage. By using this strategy, at initial ambient temperature of 5 °C, 15 °C and 25 °C, different black body temperatures were measured and listed in table 1 to table 3.

It should be noted that these experimental results in the table 1 to table 3 are obtained by calculating the output voltage of thermocouple after heating for 2 seconds, and the three temperature values recorded under the same experimental conditions are the standard black body temperatures measured at the 3$^{rd}$, 4$^{th}$ and 5$^{th}$ seconds after power on.

**Table 1** Measurement results at 5 °C laboratory ambient temperature

| Black body temperature/°C | Measured temperature/°C | | |
|---|---|---|---|
| 35 | 34.9 | 35.1 | 35.1 |
| 36 | 36.2 | 36.0 | 36.2 |
| 37 | 37.2 | 37.1 | 37.0 |
| 38 | 38.2 | 38.0 | 38.1 |
| 39 | 39.1 | 38.9 | 39.0 |

**Table 2** Measurement results at 15 °C laboratory ambient temperature

| Black body temperature/°C | Measured temperature/°C | | |
|---|---|---|---|
| 35 | 35.2 | 35.2 | 35.0 |
| 36 | 36.2 | 35.9 | 36.1 |
| 37 | 37.2 | 37.0 | 37.1 |
| 38 | 38.1 | 38.1 | 37.9 |
| 39 | 39.2 | 39.0 | 38.9 |

**Table 3** Measurement results at 25 °C laboratory ambient temperature

| Black body temperature/°C | Measured temperature/°C | | |
|---|---|---|---|
| 35 | 35.1 | 35.0 | 34.9 |
| 36 | 36.2 | 35.9 | 36.0 |
| 37 | 37.1 | 37.0 | 37.0 |
| 38 | 38.0 | 38.0 | 37.8 |
| 39 | 39.1 | 39.1 | 38.9 |

By analyzing the data in the table 1 to table 3, it is not difficult to draw a conclusion that the target temperature measurement error is only $\pm$ 0.2 ℃ under different ambient temperatures by this proposed preheating and PID control method. These experimental data further show that the method can effectively overcome the influence of thermal shock and achieve the purpose of rapid and accurate measurement.

## 5. Conclusions

In this paper, a hybrid temperature compensation method for overcoming the thermal shock influence of NCIT were presented. The NCIT is easily affected by thermal shock, and infrared ear thermometer is particularly prominent when working in low temperature environments. Thermal shock affects not only the ambient temperature measurement by NTC, but also the target temperature measurement by thermopile. In order to solve these problems, the infrared sensor preheating scheme combined with digital PID control algorithm and PWM technology is proposed for improving the measurement accuracy and speed of NCIT. This method has been experimentally verified under different laboratory ambient temperatures. The experimental results show that the proposed preheating method can make the internal ambient temperature of the infrared sensor basically reach a balance within 2 s. After that, there is little heat exchange between the external environment and the sensor. In the absence of other ambient temperature compensation algorithms, the measurement error of the standard black body temperature is only $\pm$ 0.2 ℃, and the stable measurement time is reduced to about 3 s. This study would be beneficial to improve performance of NCIT, especially the infrared ear thermometer. In the next step, we will add the ambient temperature compensation algorithm based on the research of this paper to further improve the accuracy of body temperature measurement, and apply this method to the actual NCIT products.

## Acknowledgments

This work was supported by the Science and Technology Program of Hunan Province of China (No.2017SK2164 and 2019TP1014), the Scientific Research Innovation Team of Hunan Institute of Science and Technology (No.2019-TD-10), the Scientific Research Project commissioned by Dongguan Zhenhai Electronic Technology Co., Ltd. (No.JSHT-L-2020-064), the Research Project of Teaching Reform in Hunan Province of China (No. [2019] 291-630) and the Industry-University Cooperation and Collaborative Education Project of Ministry of Education of China (201902016066 and 201702071093).